# Ionic Peltier Effect in Li-Ion Electrolytes

Zhe Cheng,[*,a] Yu-Ju Huang,[*,a] Beniamin Zahiri,[a] Patrick Kwon,[a] Paul V. Braun,[a,b,c,d,e] and David G. Cahill[a,b]

ZC and YJH contributed equally.

[a]Department of Materials Science and Engineering and Materials Research Laboratory, University of Illinois at Urbana-Champaign, Urbana, IL 61801, United States.

[b]Department of Mechanical Science and Engineering, University of Illinois at Urbana-Champaign, Urbana, IL 61801, United States.

[c]Department of Chemistry, University of Illinois at Urbana-Champaign, Urbana, IL 61801, United States.

[d]Department of Chemical and Biomolecular Engineering, University of Illinois at Urbana-Champaign, Urbana, IL 61801, United States.

[e]Beckman Institute for Advanced Science and Technology, University of Illinois at Urbana-Champaign, Urbana, IL 61801, United States.

[†] Electronic Supplementary Information (ESI) available.

*Corresponding authors: zhe.cheng@pku.edu.cn; yujujh2@illinois.edu

# ABSTRACT

The coupled transport of charge and heat provide fundamental insights into the microscopic thermodynamics and kinetics of materials. We describe a sensitive ac differential resistance bridge that enables measurements of the temperature difference on two sides of a coin cell with a resolution of better than 10 µK. We use this temperature difference metrology to determine the ionic Peltier coefficients of symmetric Li-ion electrochemical cells as a function of Li salt concentration, solvent composition, electrode material, and temperature. The Peltier coefficients $\Pi$ are negative, i.e., heat flows in the direction opposite to the drift of Li ions in the applied electric field, large, $-\Pi > 30$ kJ/mol, and increase with increasing temperature at $T > 300$ K. The Peltier coefficient is approximately constant on time scales that span the characteristic time for mass diffusion across the thickness of the electrolyte, suggesting that heat of transport plays a minor role in comparison to the changes in partial molar entropy of Li at the interface between the electrode and electrolyte. Our work demonstrates a new platform for studying the non-equilibrium thermodynamics of electrochemical cells and provides a window into the transport properties of electrochemical materials through measurements of temperature differences and heat currents that complement traditional measurements of voltages and charge currents.

# INTRODUCTION

Electronic thermoelectric materials have been extensively studied over the past several decades. The Seebeck and Peltier coefficients provide unique information about the band structure and the scattering rates of charge carriers and have applications as thermometers, heat flux sensors, and solid-state heat engines that convert between thermal and electrical energy [1,2]. Thermoelectric effects in ionic conductors also have a long history. The ionic Seebeck effect was discovered in the late 19th century [3] and the ionic ago [4]. The ionic Seebeck effect has been widely studied as a means of gaining insight into the fundamental thermodynamics and kinetics of electrochemical materials [5–7] and for potential applications in sensing and harvesting thermal energy [8–14].

In comparison to the ionic Seebeck effect, the ionic Peltier effect is relatively unexplored due to experimental challenges. First, many ionic conductors are air-sensitive. Hermetic sealing of ionic conductors adds additional thermal resistance to the system which complicates the measurements. Second, the thermoelectric figure of merit of ionic conductors is small and limits the size of the temperature differences that can be generated by a charge current. In other words, the small currents that are needed to avoid significant Joule heating produce small Peltier effect was first accurately measured nearly a century temperature differences at the two ends of an ionic conductor. This problem of electrical dissipation is aggravated by large electrical resistances that are commonly observed at interfaces between electrodes and electrolytes. Third, even if Joule heat can be accurately separated from Peltier heat at high current densities, current densities that can be applied to an electrochemical cell are often constrained by instabilities, e.g., the formation of dendrites.

Schmid and co-workers [15,16] overcame many of these challenges using a microcalorimeter with a fast response time and 10 ms pulses of electrical currents that localize the heat near the sensor. These experiments probed the fast transient response of the electrode but could not access time scales where the electrochemical cell reaches a steady state. In our work, we take a different approach that allows us to span a wide range of time scales and measure both the transient and steady-state behavior. We seal liquid electrolytes in a thin, symmetric coin cell with Li metal or partially delithiated $Li_{1-x}CoO_2$ electrodes. The standard coin cell geometry increases the throughput of our work. We implement a temperature difference metrology (TDM) with a sensitivity of 10 µK using a differential ac resistance bridge and two thermistors attached on each side of the coin cell.

This TDM and the symmetric cell design enable the measurements of the reversible ionic Peltier heat while minimizing the effect of irreversible Joule heat. In a perfectly symmetric cell, Joule heat raises the average temperature of the cell but does not produce a temperature difference. We interpret our data for the Peltier coefficient in terms of the changes in entropy produced by the reactions at the electrodes, and the heat of transport of the ionic species. We find that the absolute magnitude of the Peltier coefficient of the electrode/electrolyte interfaces is large, greater than 30 kJ/mol, and negative in sign, i.e., the heat current is in the direction opposite to the charge current.

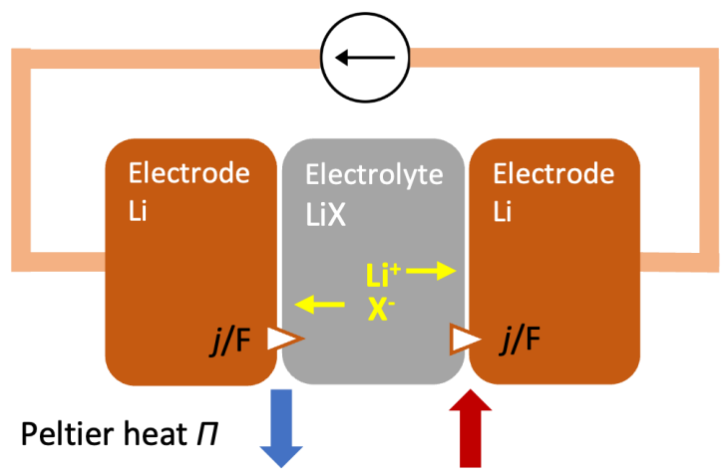

Fig. 1 Schematic illustration of the symmetric Li ion electrochemical cell. The electrodes are either Li metal or $Li_{0.75}CoO_2$. The electrolyte is either lithium bis(trifluoromethanesulfonyl)imide (LiTFSI) dissolved in 1,2-dimethoxyethane (DME) or $LiPF_6$ dissolved in a 50/50 (v/v) mixture of ethylene carbonate (EC) and dimethyl carbonate (DMC). The Li salt is denoted LiX.

## BACKGROUND

Non-equilibrium thermodynamics (NET) provides a theoretical framework for understanding the relationships between transport coefficients. The NET of coupled mass, heat, and charge transport in electrochemical cells subjected to a temperature gradient, so-called thermogalvanic cells, has been thoroughly described in a book chapter by Agar [17]. Kjelstrup and co-workers recently provided a modern perspective on the NET of thermogalvanic cells [18], building on their prior work [19–23]. Our Fig. 1 is adapted from Fig. 1 of ref. 18 and modified to depict the symmetric Li ion electrochemical cells that we studied. Li cations form at the anode/electrolyte interface and are consumed at the cathode/electrolyte interface. The anion, either bis(trifluoromethanesulfonyl)imide or PF6⁻, does not participate in redox reactions at the electrodes and is denoted X⁻ in Fig. 1 and the discussion below.

Following ref.18, we define the Seebeck coefficient and Peltier coefficients of the thermogalvanic cell as

$$\eta_s = \frac{\Delta \varphi}{\Delta T} \quad (1)$$

$$\Pi = \frac{F J_q}{j} \quad (2)$$

where $\varphi$ is the electric potential, $J_q$ the heat current, $j$ the charge current, and $F$ is the Faraday constant. (This definition of the Seebeck coefficient, Eq. 1, is common in the physical chemistry community but has a different sign from the definition that is common for electronic materials. For metals and semiconductors, the Seebeck coefficient is usually defined as the electric field divided by the temperature gradient.) The Seebeck and Peltier coefficients are related by an Onsager reciprocal relationship

$$\eta_s F T = -\Pi \quad (3)$$

A positive current is defined by a flow of positive charge through the cell from left to right. We accordingly label the left electrode the anode (a) and the right electrode the cathode (c) since a positive current produces oxidation at the left electrode and reduction at the right electrode. The voltage and temperature differences are defined from right to left, i.e., $\Delta \phi = \phi_c - \phi_a$ and $\Delta T = T_c - T_a$.

The depiction of the Peltier heat in Fig. 1 is for the case when the Peltier coefficient of the electrolyte is more positive than the Peltier coefficient of the electrodes: heat transfers from the thermal bath to the anode/electrolyte interface and heat transfers from the cathode/electrolyte interface to the thermal bath. In fact, we find in our experiments that the Peltier heat has a negative sign, i.e., the opposite sense of what is shown in Fig. 1, and therefore the Peltier coefficient of the electrolyte is more negative than the electrode.

As discussed in Ref.18, to analyze the coupled transport of heat, mass, and charge, we must choose components, control variables, and a frame of reference. Values of the heat of transport depend on the frame of reference. Choosing neutral components and one of the solvent components as the frame of reference is the most natural and facilitates comparisons between experiment and theory [18].

The Peltier coefficient is expected to depend on the time after the start of an electrical current. On short time scales, short compared to the characteristic time scale for diffusion of mass in the electrolyte $\tau_D = h^2/(\pi^2 D)$ where h is the electrolyte thickness, the composition of the electrolyte is uniform and the chemical potential of the components is constant. In this situation, the Peltier coefficient is[20]

$$\frac{\Pi}{T} = S^*_{Li^+} - S_{Li} - \frac{(t_{X^-})q^*_{LiX}}{T} \quad (4)$$

where $S_{Li}$ is the partial molar entropy of Li in the electrode, $S^*_{Li^+}$ is the transported entropy of the Li ion, $t_{X^-}$ the transport coefficient for the anion, and $q^*_{LiX}$ the heat of transport of the LiX neutral component. (For a binary solvent, e.g., a mixture of EC and DMC, there is an additional term for the heat of transport of the solvent component that is not the frame of reference [20]. We are omitting that term here because we have not observed transient effects in the heat currents and the added complexity does not seem merited.)

We omitted the electronic transported entropy in the electrode in the expression above because that term is negligible in comparison to the large Peltier effects we observe in our experiments. In our experiment, the electrodes are electrically connected to stainless steel plates and the stain- less steel case of the coin cell and then to Cu wires. Since the thermal conductivity of the electrodes and stainless steel is large, the effects of electronic transported entropy cancel out except for the

Cu leads.

On time scales long compared to τ_D, there is a concentration gradient of the anion X⁻ such that the diffusion and drift currents are equal and opposite [24]. Charge neutrality requires an equivalent concentration gradient of Li⁺ and therefore there is also an equivalent concentration gradient in the neutral component LiX. In a steady state, the only mass and charge transport in the electrolyte is the Li ions. The Peltier coefficient of the cell is then [20]

$$\frac{\Pi}{T} = S^*_{Li^+} - S_{Li} \quad (5)$$

The transported entropy is the sum of the partial molar entropy and the heat of transport divided by temperature. (The heat of transport divided by temperature is synonymous with the Eastman entropy of transport.)

$$S^*_{Li^+} = S_{Li^+} + \frac{q^*_{Li^+}}{T} \quad (6)$$

Therefore, in this limit of steady-state and a single component solvent,

$$\Pi = T\,(S_{Li^+} - S_{Li}) + q^*_{Li^+} \quad (7)$$

The first term, $T\,(S_{Li^+} - S_{Li})$, is a thermodynamic property of the system at equilibrium, the change in the partial molar entropy between Li in the electrode and Li⁺ in the electrolyte. The changes in the entropy of the solvent that are produced by complexes of Li⁺ and organic solvent molecules have been studied by many authors with the goal of better understanding the thermodynamics of ion solvation [7]. Solvated Li ions are thought to strongly reduce the entropy of the solvent by restricting the motion and orientations of the solvent molecules within the first coordination shell. Schmid et al. [15] used microcalorimetry data and analysis of the change in entropy to estimate the number of solvent molecules within the first coordination shell.

The second term, $q^*_{Li^+}$, the heat of transport of Li⁺, is a transport coefficient. (This term is often neglected in the analysis of the Seebeck coefficient of thermogalvanic cells [5,7]). Non-equilibrium thermodynamics (NET) pro- vides a rigorous description of the relationships between transport coefficients and enables the interpretation of experiments in terms of those transport coefficients. NET cannot, however, provide the values of the transport coefficients. A microscopic kinetic theory is needed to predict the values of the transport coefficients from the composition and structure of a material. There have been many attempts to describe the heat of transport of liquid [25] and solid [26] electrolytes at a microscopic level. A comprehensive theory of the heat of transport of ions remains elusive to the present day.

Würger [27] recently revisited kinetic models [28] for hop- ping transport and deduced from a simple model that the heat of transport is given by $q^* = RT + \Delta H$ where $\Delta H$ is the change in enthalpy at the transition state that separates two equilibrium configurations of the ion. Würger's model is based on local thermal equilibrium and does not include effects that involve, e.g., fluctuations of atomic configurations and vibrational occupations that are not described by an effective temperature.

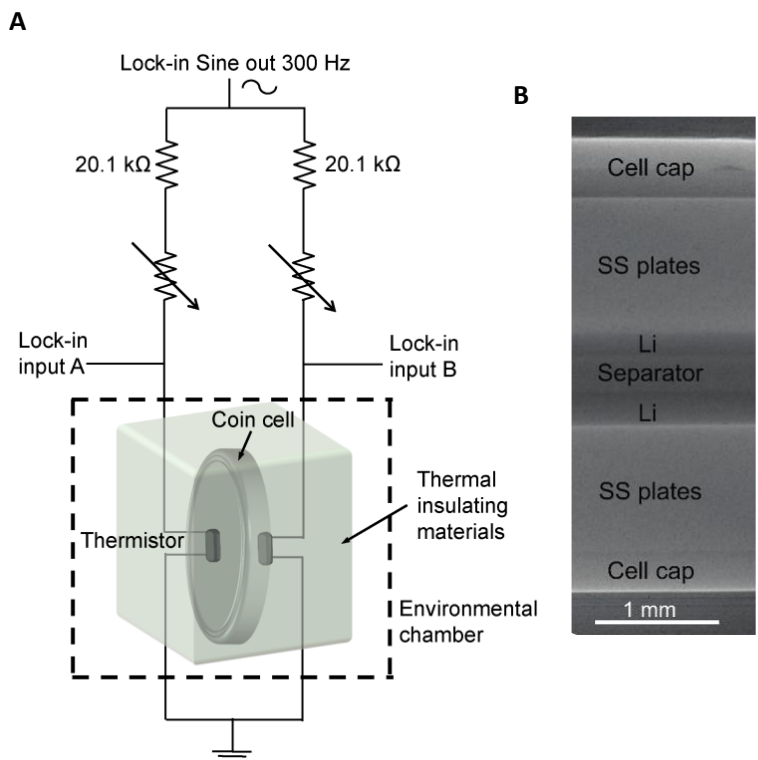

Fig. 2 (A) Schematic diagram of the measurement of the ionic Peltier coefficient. Two thermistors

are attached on opposite sides of a symmetric coin cell. The coin cell which is surrounded by thermally insulating materials (polymer foam) is placed in an environmental chamber. An ac bridge is used to measure the difference between the electrical resistance of the two thermistors and record the temperature difference of the two sides of the coin cell. (B) Computed tomography (CT) cross-sectional image of a symmetric coin cell. The orientation of the CT image is rotated by 90 degrees relative to the schematic in panel (A). Here lithium metal is used as the electrodes. Two stainless steel (SS) plates are used as spacers on each side of the lithium metal electrode.

## RESULTS

Electrodes and liquid electrolytes are sealed in a symmetric coin cell, as shown in Fig. 2. Two thermistors are adhered to the surface of the two sides of the symmetric coin cell by cyanoacrylate adhesive. The coin cell is surrounded by thermally insulating materials (porous polymer foam) and placed in an environmental chamber (ESPEC SH242).

We use an electrical bridge with an ac excitation volt- age to measure the difference in electrical resistance of two thermistors [29] corresponding to the difference in temperature at the two sides of the coin cell. The metal-oxide thermistors have a volume of ∼ 1.0 mm³ and a 10 kΩ resistance at room temperature. The temperature difference ΔT is derived from the out-of-balance voltage of the bridge ΔV using

$$\Delta T = -\frac{\Delta V}{V_0}\left(\frac{dR}{dT}\right)^{-1}\frac{(R_s+R)^2}{R_s} \quad (8)$$

where $V_0$ is the excitation voltage of the bridge, R is the average resistance of the thermistors, and $R_s$ is the resistance of the series resistors. To a good approximation, the temperature dependence of the thermistor resistance is $dR/dT = -RB/T^2$, where the so-called B parameter of the thermistors we are using is B ≈ 3400 K. Therefore, near room temperature the normalized temperature coefficient is large compared to unity: $(dR/dT)(T/R) = -B/T \approx 10$. The bridge is optimized with $R_s \approx R$. The noise of the relative temperature measurement (ΔT/T) is then a few times smaller than the noise of the relative voltage measurement (ΔV/$V_0$). Since the thermal noise generated by a single 10 kΩ resistor is 13 nV/√Hz, with $V_0$ = 0.5 V, the fundamental limit on the noise in the temperature difference measurement is approximately 5 μK/√Hz.

Fig. 2B shows a computed tomography (CT) cross-section of the symmetric coin cell. Heat absorption or release due to the ionic Peltier effect occurs at the electrode (Li metal or delithiated $Li_{1-x}CoO_2$)-electrolyte interfaces. In steady-state, the Peltier heat is compensated by heat flow through the separator/liquid-electrolyte composite. The thermal conductance of the separator/liquid- electrolyte composite (G) is much larger than the thermal conductance of natural convection from cell cap to air. As a result, the heat flux through the Li, stainless- steel spacers, cell cap, and adhesive layers is negligible. Thus, the difference between temperatures measured by the thermistors and the temperatures of the two sides of the separator/liquid-electrolyte composite is negligible. We estimate < 2% systematic error in the measurement of the temperature difference coming from heat losses to the surroundings.

We determine the ionic Peltier heat current from the temperature difference using $J_q = \Delta T G$. The sign convention in this equation corresponds to Fig. 1. A positive Peltier heat creates a positive ΔT. The thickness of the separator/liquid-electrolyte composite is determined by x- ray computed tomography. The thermal conductance G of the electrolyte/separator composite is calculated by effective medium theory [30]. Additionally, the perimeter rubber O-ring which is used to seal the coin cell also contributes to the thermal conductance. We estimate the thermal conductance of the O-ring to be approximately 10% of the thermal conductance of the separator/liquid-electrolyte composite. More details can be found in the Supporting Information (SI) Appendix.

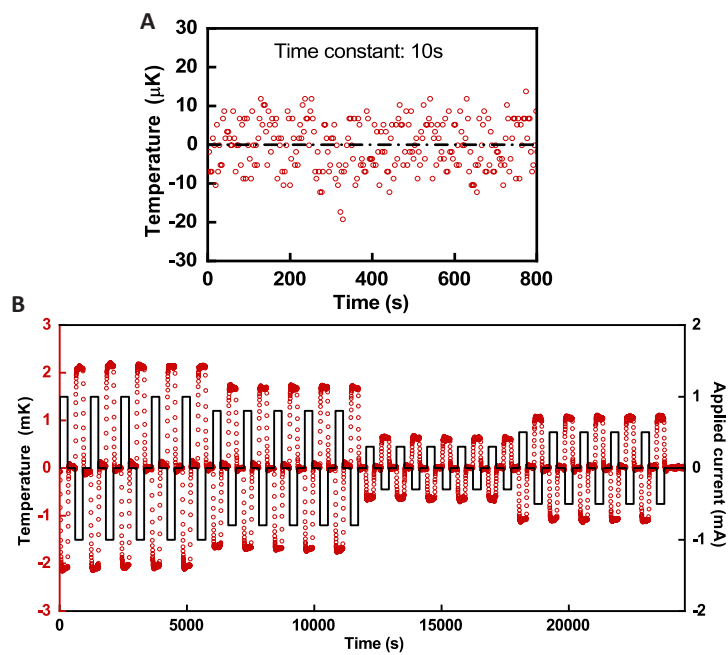

Fig. 3 Examples of temperature difference metrology (TDM). (A) Resolution of TDM with a lock-in time constant of 10 s. (B) Temperature difference response of the two sides of the symmetric coin cell containing 2 mol L$^{-1}$ LiPF$_6$-EC-DMC when applying electrical currents to the coin cell. The temperature difference is plotted as red symbols versus the left axis; the applied current is plotted as a black line against the right axis. The area of the electrodes is 1.90 cm$^2$.

The resolution of the temperature difference metrology (TDM) is illustrated in Fig. 3A. The peak-to-peak fluctuations are 20 µK with the time constant of the lock-in amplifier set to 10 s. One data point is recorded every 3 s. We use a 10 s time constant in all of the following ionic Peltier heat measurements.

The temperature differences between the two sides of a coin cell containing 2 mol L$^{-1}$ LiPF$_6$-ethylene carbonate (EC)-dimethyl carbonate (DMC) liquid electrolyte generated by applying an electrical current are shown in Fig. 3B. (LiPF$_6$-EC-DMC is commonly used as the electrolyte in Li-ion batteries.) When the sign of the current changes, the sign of the temperature difference also changes. The parasitic Joule heat in the coin cell and the ac excitation cur- rent induced Joule heat in the thermistors affect the over- all temperature rises of the coin cell and the thermistors (∼ 300 mK estimated as the Joule heating power divided by thermal conductance created by natural air convection and radiation to the environment) but does not affect the ionic Peltier heat measurements. The symmetric design of the experiment suppresses contributions to the temperature difference that are produced by terms that are even in the current.

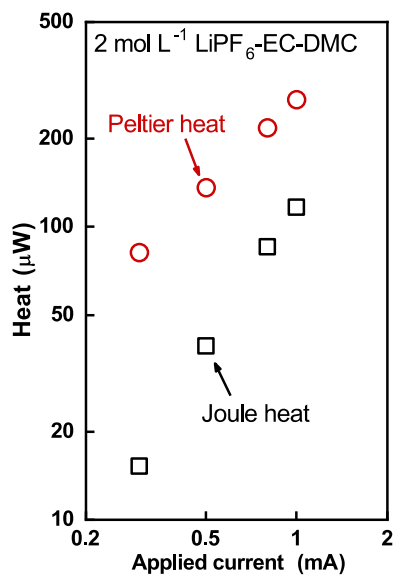

Fig. 4 The ionic Peltier heat power and Joule heat power of 2 mol L$^{-1}$ LiPF$_6$-EC-DMC as a function of applied currents.

The corresponding ionic Peltier heat and Joule heat per unit time of 2 mol L$^{-1}$ LiPF$_6$-EC-DMC are plotted in Fig. 4 as a function of applied current. The Joule heat generation rate is calculated as voltage times current. The ionic Peltier heat per unit time is derived from the temperature difference $\Delta T$ and varies linearly with the current. In this set of measurements, the ionic Peltier heat is larger than the Joule heat because the internal resistance of the cell is relatively small at room temperature (110 to 160 $\Omega$, de- pending on applied currents).

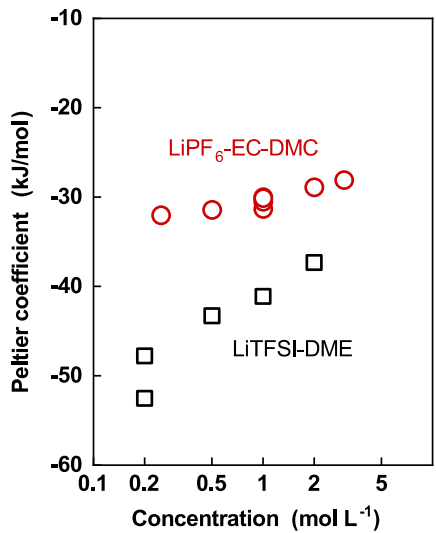

Fig. 5 Ionic Peltier coefficients of LiPF$_6$-EC-DMC and LiTFSI- DME with different salt concentrations.

We further performed measurements on different salt concentrations for both LiPF$_6$-EC-DMC and LiTFSI-DME (DME is 1,2-dimethoxyethane) electrolytes; the results are summarized in Fig. 5. The ionic Peltier coefficient becomes less negative with increasing salt concentrations. The dependence on salt concentrations for the LiTFSI-DME electrolyte is stronger than for the LiPF$_6$-EC-DMC electrolyte. The absolute magnitudes of the ionic Peltier coefficients are approximately one order of magnitude higher than typical electronic thermoelectric materials [27,31] and similar to the Peltier coefficient of lightly doped Si [32].

The Peltier coefficients that we measured are compared with Peltier coefficients derived using the Onsager relation- ship (Eq. 3) from recently published measurements of the electrode potential temperature coefficients of closely related electrolytes (LiTFSI-DOL-DME and LiPF$_6$-EC-DEC) [7], see Fig. S5 in the SI Appendix. Our measurements also agree with prior work on the Seebeck coefficients of symmetric thermogalvanic cells that was reviewed by Gunnarshaug and co-workers [23].

We also checked the effect of electrode composition on the ionic Peltier coefficients. We found that the ionic Peltier coefficients of 1 mol L$^{-1}$ LiPF$_6$-EC-DMC using Li metal electrodes and

Li$_{0.75}$CoO$_2$ electrodes are the same ($\Pi = -30$ kJ/mol) within our experimental uncertainties, consistent with previous measurements on electrode potential temperature coefficient as a function of electrode composition [7]. We used Li metal electrodes in all other measurements.

We emphasize that we use electrodes that are redox active for Li and thereby directly connect the transport of Li$^+$ in the electrolyte to external electronics for the control of currents and voltages. The mechanisms of thermogalvanic cells that use redox-active electrodes are different from the recently observed giant ionic Seebeck coefficients which typically used blocking electrodes and experimental designs based on capacitors or supercapacitors [11–13, 33]. With the use of blocking electrodes, the measurements must be performed in a transient manner because the impedance of the device is extremely large in the limit of long time or low frequency.

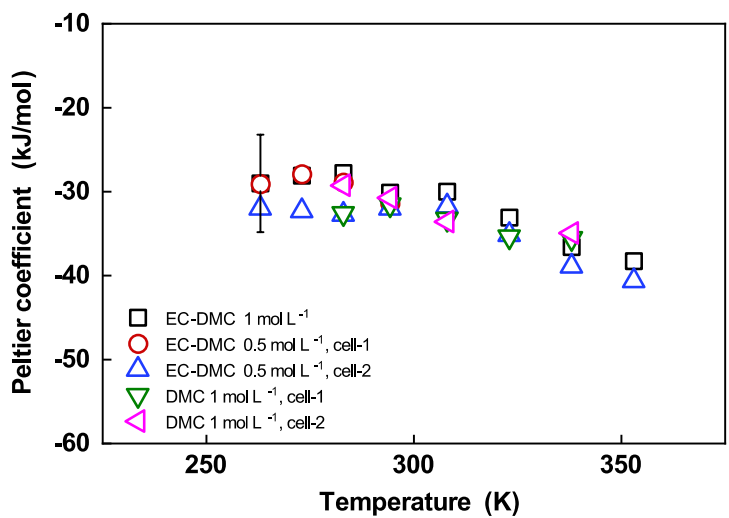

Fig. 6 Temperature-dependent ionic Peltier coefficients of 1 mol L$^{-1}$ LiPF$_6$-DMC and LiPF$_6$-EC-DMC for 0.5 mol L$^{-1}$ and 1 mol L$^{-1}$. Cell-1 and cell-2 are two cells with the same compositions.

To further understand the ionic Peltier effect, we measured the temperature dependence of the Peltier coefficients by placing the experiment within a temperature-controlled environmental chamber. The temperature dependent ionic Peltier coefficients of 1 mol L$^{-1}$ LiPF$_6$-DMC and LiPF$_6$-EC-DMC for 0.5 mol L$^{-1}$ and 1 mol L$^{-1}$ are shown in Fig. 6. We are not aware of any previous reports of the temperature dependence of the Seebeck of Peltier coefficients in a thermogalvanic cell. The measured temperature range is limited by the low ionic conductivity of liquid electrolytes and SEI layers at low temperatures, and the thermal stability of liquid electrolytes at high temperatures. The results show that the ionic Peltier coefficient is approximately constant below room temperature but in- creases with increasing temperature at T > 300 K. The increase in ionic Peltier coefficients at higher temperatures is observed for both the 0.5 mol L$^{-1}$ and 1 mol L$^{-1}$ LiPF$_6$-EC-DMC electrolytes. (The coin cell (cell-2) with 0.5 mol L$^{-1}$ LiPF$_6$-EC-DMC shorted out at low temperatures so only four data points are measured and included here.) The temperature dependences of ionic Peltier coefficients of 1 mol L$^{-1}$ LiPF$_6$-EC-DMC and 1 mol L$^{-1}$ LiPF$_6$-DMC are similar. At T > 300 K, the Peltier coefficient is consistent with a constant value of Π/T.

## DISCUSSION

As summarized above in the Background section, the ionic Peltier effect in a symmetric cell has two contributions: the difference in the equilibrium partial molar entropy of Li in the electrode and electrolyte T ($S_{Li+} - S_{Li}$), and the heat of transport of Li in the electrolyte q*$_{Li+}$ [17,18,20].

We now consider the sign and size of these terms. The change in the partial molar entropy $S_{Li+}$ has contributions from the configurations of Li$^+$ in the electrolyte, the alignment and restricted motion of solvent molecules in the electrolyte by interactions with Li$^+$, and changes in vibrational frequencies of the Li atom and the surrounding sol- vent molecules. The Li$^+$ configurational entropy is positive and the change in entropy created by the alignment of solvent molecules is expected to be strongly negative [16,34]. Since each Li$^+$+ ion affects the alignment of several solvent molecules, the partial molar entropy of Li$^+$ in the electrolyte can be large.

It is known that Li$^+$ solvation will introduce changes in the ordering of solvent molecules by the influence of the electric field [35] of the cation. The entropy decreases upon crystallization of a liquid solvent, i.e., the entropy of fusion, has often been considered as a benchmark for the loss of solvent entropy due to ion solvation. If the Li$^+$-solvent interaction reduces the configurational entropy of the solvent molecules in the primary solvation shell to a degree that is comparable to the crystallization of the solvent, then the maximum entropy decrease caused by the Li$^+$-solvent interaction will be NΔS$_{fus}$, where N is the number of solvent molecules in the first solvation shell.

In our experiments, if we assume that the heat of transport is small, we observe a change in the partial molar enthalpy comparable to 100 J/(mole-K). A typical entropy of fusion for a carbonate solvent is 40 J/(mole-K). Since N ≈ 4, we observe a change in partial molar entropy that is ≈ 60% of the limit predicted by the entropy of fusion.

We note that the entropy of solvation has often been re- ported to be a smaller fraction of $N\Delta S_{fus}$ than what we derive from our work. For example, the entropy of fusion of DMSO [36] is -48 J/(mol-K); and the entropy of $Li^+$ solvation in DMSO [35,37] is -59 J/(mole-K). Thus, for DMSO, and assuming N = 4, the entropy of solvation is ≈ 30% of the limit predicted by the entropy of fusion.

In the discussion above, we have made the assumption that the contribution of the $Li^+$ heat of transport to the Peltier coefficient is small compared to the change in partial molar entropy. The strongest evidence to support this assumption is the lack of time dependence of the Peltier coefficient. The thickness of the electrolyte is on the order of 300 < h < 500 μm. The small thickness of the electrolytes allows us to study both the short-time (Eq. 4) and steady-state Peltier coefficients (Eq. 5). Diffusion in the electrolyte will be limited by the component of the electrolyte with the smallest diffusion constant, in this case, Li. We estimate that the diffusion constant of Li is in the range of $1 - 3 \times 10^{-10}$ m²/s and therefore the time-scale for mass transport in the electrolyte is $\tau = h^2/(\pi^2 D)$, 30 < τ < 250s. We have not observed a significant change in the Peltier coefficient on these time scales and we conservatively set an upper bound of 3 kJ/mol for the difference between the short-time and steady-state values of the Peltier coefficient. As discussed in the Background section, this difference is expected to be $(t_{X-})q^*_{LiX}$. The heat of transport of the neutral component LiX is the sum of the heat of transport of the ions 18, $q^*_{LiX} = q^*_{Li+} + q^*_{X-}$. Unless the two ionic heats of transport have approximately equal magnitude and opposite sign, we conclude that the absolute magnitude of the heat of transport of $Li^+$ makes a minor contribution to the Peltier coefficient.

In kinetic models for the heat of transport, the heat of transport is often related to the activation energy for diffusion but can have either a positive or negative sign depending on if the activation energy is associated with fluctuations of the ion or associated with fluctuations of the solvent. The $Li^+$ self-diffusion coefficient in organic electrolytes has been measured by the pulse-gradient spin-echo nuclear magnetic resonance experiments from 289 K to 353 K [38]. The activation energy derived from the temperature dependence of the self-diffusion coefficient of Li ions in 1 mol L$^{-1}$ LiPF$_6$-EC-DMC is 13 kJ mol−1 [38]. The absolute magnitude of the ionic Peltier coefficients we observe in our work is a factor of ∼ 3 larger than what is expected from the heat of transport alone.

Finally, we comment on the observed decrease in the absolute magnitude of the Peltier coefficient with increasing salt concentration, a result that is counter to expectations based on the configurational entropy of Li$^+$. Assuming that the Peltier coefficient is dominated by the difference between the partial molar entropy of Li+ in the electrolyte and Li in the electrode, a high salt concentration would have a lower Li$^+$ configurational entropy, and therefore the entropy change would be more negative at higher salt concentrations. Our data show that the Peltier coefficient is less negative at higher salt concentrations. Interactions between Li$^+$-solvent complexes play a more important role in determining the dependence of the Peltier coefficient on salt concentration than the Li$^+$ configurational entropy [16].

# CONCLUSIONS

In conclusion, we report measurements of the ionic Peltier effect by a temperature-difference-metrology (TDM). We observe large, negative values of the ionic Peltier coefficients in LiPF$_6$-EC-DMC and LiTFSI-DME liquid electrolytes over a range of salt (LiPF$_6$ or LiTFSI) concentrations. The ionic Peltier coefficients become less negative with increasing salt concentrations. By performing temperature-dependent ionic Peltier heat measurements, we found that the measured ionic Peltier coefficients are weakly temperature-dependent below room temperature and become more negative with increasing temperature above room temperature. The lack of significant time- dependence in the Peltier coefficients suggests that the heat of transport of the ionic species plays only a minor role and that the Peltier coefficient is dominated by the difference in partial molar entropy of the electrode and electrolyte.

# MATERIALS AND METHODS

**Fabrication of symmetric coin cells**

The coin cells in this work are symmetric cells containing the liquid electrolytes $LiPF_6$-EC-DMC 50/50 (v/v) and LiTFSI-DME. $LiPF_6$-EC-DMC is purchased from Sigma Aldrich while the LiTFSI-DME is mixed in a glovebox in our laboratory. DME was purchased from Sigma Aldrich and LiTFSI salt was procured from Tokyo Chemical Indus- try. One test cell used electrodes of commercially prepared $LiCoO_2$ on Al foil (MTI) at a state-of-charge of approximately 50% corresponding to $Li_{0.75}CoO_2$; all other cells used Li metal as electrodes. The thickness and diameter of the Li metal electrodes purchased from MTI are 250-300 µm and 15.6 mm. The separators were one or two layers of grade GF/D glass microfiber purchased from Whatman GE Healthcare Life Sciences. More details can be found in the SI Appendix. Stainless steel 304 spacers and 2032 coin cell parts are purchased from MTI. The coin cell was assembled by placing two 0.5 mm thick stainless steel spacers on one coin cell cap, followed by the first electrode (Li metal or $LiCoO_2$), 2 separators (a few drops of electrolytes applied to fully wet the separators), Li metal, two 0.5 mm thick stainless steel spacers, followed by the second coin cell cap. All are pressed together at a pressure of 0.8 metric tons.

**Temperature difference metrology**

Thermistors (1.6 mm long, 0.8 mm wide, and 0.8 mm thick) are purchased from Digi-Key (Murata, NCP18XH103D03RB, 10 kΩ at room temperature). The resistive wire used to connect the thermistors is nichrome 80 with a diameter of 0.13 mm and a resistance of 120 Ω m$^{-1}$. To keep good thermal contact and reliable electrical insulation between the thermistors and the cell surface, cyanoacrylate adhesive is used to adhere the thermistors on the surface of the two sides of the coin cell. After the cyanoacrylate adhesive is dry, epoxy adhesive is added to cover the thermistors to further fix the thermistors. The coin cell is connected to a DC current source (Lake Shore 121) by thin copper wires adhered by silver paste [29]. A lock-in amplifier (Stanford Research SR830) is used to excite the electrical bridge (300 Hz, 0.5 V) and detect the voltage difference of the two arms of the bridge, as shown in Fig. 2A. More details can be found in the SI Appendix. The coin cell is put inside an environmental chamber (ES-PEC SH242) to control the environmental temperature. Because there is a fan inside the environmental chamber, the coin cell is surrounded by porous polymer foam to retard any forced convection heat transfer. The variation in environmental temperature is controlled to be less than ±0.1 K. We waited more than 4 hours after changing each chamber temperature point to ensure the environmental temperature reaches a steady state. Room temperature in this work is 22 °C determined by the resistance of the thermistor.

**Computed tomography (CT)**

Images of the cross-section of the coin cells are measured by either an Xradia micro-CT or a Rigaku CTLab GX130 at the Beckman Institute for Advanced Science and Technology at the University of Illinois. The thickness of the separator/liquid-electrolyte composite is derived from the computed tomography images.

**Data availability**

Data for the geometries of the coin cells and the Peltier coefficients derived from the temperature-difference measurements are tabulated in the Supporting Information (SI) Appendix.

## CONFLICTS OF INTEREST

There are no conflicts to declare.

## ACKNOWLEDGEMENTS

The authors acknowledge the helpful discussion with Profs. Alois Würger and Signe Kjelstrup. We thank Arghya Patra for fabricating part of the coin cells and Peilin Lu for synthesizing the LiTFSI-DME electrolyte. The authors disclose financial support for the research of this work before September 2023 from US Army CERL W9132T-19-2-0008 and US Army CERL W9132T-21-2-0008. The experimental results produced after September 2023 by Yu-Ju Huang for the changes in the Peltier coefficient as a function of temperature and composition, and the writing of the final version of the paper by Yu-Ju Huang and David Cahill were supported by the U.S. Department of Energy, Office of Basic Energy Sciences, Division of Materials Sciences and Engineering under Award Number DE-SC0024461.